\documentclass[useAMS]{mn2e}
\usepackage{graphicx}  
\usepackage{subfig}
\usepackage{amssymb}
\usepackage{natbib}
\usepackage{amsmath}

\DeclareGraphicsExtensions{.ps,.pdf,.png}
\makeatletter
\def\fps@figure{htbp}
\makeatother

\begin{document}

\title[Brighter galaxy bias]{Brighter galaxy bias: underestimating the velocity dispersions of galaxy clusters}

\author[Old et al.]{L. Old$^{1}$\thanks{E-mail: ppxlo@nottingham.ac.uk}, M. E. Gray$^{1}$ and F. R. Pearce$^{1}$ \\
$^{1}$University of Nottingham, School of Physics and Astronomy, Nottingham, NG7 2RD UK}

\date{Accepted ??. Received ??; in original form ??}
\pagerange{\pageref{firstpage}--\pageref{lastpage}} \pubyear{2011}
\maketitle

\label{firstpage}

\begin{abstract}
We study the systematic bias introduced when selecting the spectroscopic redshifts of brighter cluster galaxies to estimate the velocity dispersion of galaxy clusters from both simulated and observational galaxy catalogues. We select clusters with $N_{\rm gal} \geq 50$ at five low redshift snapshots from the publicly available De Lucia \& Blaziot semi-analytic model galaxy catalogue. Clusters are also selected from the Tempel SDSS DR8 groups and clusters catalogue across the redshift range 0.021 $\leq z \leq$ 0.098. We employ various selection techniques to explore whether the velocity dispersion bias is simply due to a lack of dynamical information or is the result of an underlying physical process occurring in the cluster, for example, dynamical friction experienced by the brighter cluster members. The velocity dispersions of the parent dark matter (DM) halos are compared to the galaxy cluster dispersions and the stacked distribution of DM particle velocities are examined alongside the corresponding galaxy velocity distribution. We find a clear bias between the halo and the semi-analytic galaxy cluster velocity dispersion on the order of $\sigma_{\rm gal} / \sigma_{\rm DM} \sim 0.87-0.95$ and a distinct difference in the stacked galaxy and DM particle velocities distribution. We identify a systematic underestimation of the velocity dispersions when imposing increasing absolute I-band magnitude limits. This underestimation is enhanced when using only the brighter cluster members for dynamical analysis on the order of $5-35\%$, indicating that dynamical friction is a serious source of bias when using galaxy velocities as tracers of the underlying gravitational potential. In contrast to the literature we find that the resulting bias is not only halo mass-dependent but that the nature of the dependence changes according to the galaxy selection strategy. We make a recommendation that, in the realistic case of limited availability of spectral observations, a strictly magnitude-limited sample should be avoided to ensure an unbiased estimate of the velocity dispersion.\\
\end{abstract}

\begin{keywords}
galaxies: clusters - cosmology: observations - galaxies: haloes - galaxies: kinematics and dynamics
- methods: numerical

\end{keywords}


\section{Introduction}
The study of the origin, evolution and eventual fate of the Universe remains at the frontier of modern astronomy. Playing a key role in physical cosmology, galaxy clusters lie at the focus of current and impending surveys. With typical masses in the range of $10^{14} - 10^{15} M_{\odot}$, these agglomerates of galaxies are the largest gravitationally bound structures in the Universe, producing observable signatures across the electromagnetic spectrum. Clusters are most commonly detected via overdensities in the number counts of galaxies (e.g., \citealt{1958ApJS....3..211A}, \citealt{1968cgcg.book.....Z}), as overdensities of red galaxies in both the optical and infra-red (e.g., \citealt{2005yCat..21570001G}, \citealt{2007ApJ...660..221K}, \citealt{2011ApJ...736...21S}, \citealt{2012MNRAS.420.1167A}), as X-ray bright extended sources (e.g., \citealt{1972ApJ...178..309F}, \citealt{2000ApJS..129..435B}, \citealt{2002ARA&A..40..539R}, \citealt{2009ApJ...692.1033V}), as Cosmic Microwave Background distortions (e.g., \citealt{1972CoASP...4..173S}, \citealt{2002ARA&A..40..643C}, \citealt{2013arXiv1303.5089P}, \citealt{2010ApJ...722.1180V}) and as the source of weak lensing shear of background galaxies (e.g., \citealt{1990ApJ...349L...1T}, \citealt{2001PhR...340..291B}). Despite the diverse methods with which to probe clusters, obtaining accurate cluster masses remains a nontrivial problem that limits the constraining power of clusters for cosmological parameter estimation (\citealt{2011ARA&A..49..409A}). The uncertainties in cluster mass measurement are due to the fact that clusters are not always simple, spherical, virialised objects; they are extremely complex, evolving systems with unusual features often only visible at certain wavelengths.\\
\indent Dynamical analysis of clusters using spectroscopic redshifts of member galaxies exposes substructure in the form of asymmetrical velocity distributions and dynamically distinct subgroups (e.g., \citealt{1982PASP...94..421G}, \citealt{1988AJ.....95..985D}, \citealt{2012MNRAS.421.3594H}, \citealt{2012MNRAS.419.1017C}, \citealt{2012A&A...540A.123E}). Not only does this type of analysis offer key insights into the dynamical state of cluster, it also provides an estimate of cluster mass. Although velocity dispersion inferred masses are not subject to complicated baryonic physics in the intra-cluster medium that complicate X-ray mass estimates, nor the large scatter arising from the issue of deprojection in weak lensing mass estimates, velocity dispersion estimates are still prone to systematic bias. The inclusion of galaxies presumed to be cluster members that are not actually gravitationally bound to the cluster (i.e., interlopers) is a significant source of contamination (e.g., \citealt{1983MNRAS.204...33L}, \citealt{1997NewA....2..119B}, \citealt{1997ApJ...485...39C}, \citealt{2006A&A...456...23B}, \citealt{2007A&A...466..437W}). Merging activity and in-falling groups indicated by the presence of dynamical substructure invalidate the assumptions of post-virialisation (\citealt{2010MNRAS.408.1818W}). \\
\indent The idea that the galaxies are biased tracers of the gravitational potential well due to dynamical friction was first proposed by \cite{1992ApJ...396...35B}, who demonstrate that galaxies brighter than the magnitude of the third-ranked object in the cluster have velocities lower than average for a sample of 68 clusters. \cite{2005MNRAS.359.1415G} used a composite cluster of 14548 member galaxies out of 335 clusters from the SDSS DR2 to show that populations of massive galaxies have smaller velocity dispersions, postulating that this is evidence of massive galaxies losing their velocity by dynamical interaction/friction of galaxies through energy equipartition. Recent work by \cite{2012arXiv1203.5708S} identifies a clear bias introduced in the dynamical mass of semi-analytic clusters when selecting subsamples of red-luminous galaxies to estimate the velocity dispersion. They find that the impact of dynamical friction on the estimation of velocity dispersion and dynamical mass varies little with cluster mass or redshift.\\
\indent Whether it is possible to recover the true halo mass even in the case of complete member galaxy velocity information remains ambiguous, with several works finding a discrepancy between the velocity dispersion obtained using the galaxies and DM particles of up to $10\%$. Although the size of this discrepancy remains similar across the literature, there is debate as to whether the bias is positive or negative, i.e. if the velocity dispersions are, on average, larger for the DM haloes (\citealt{1996ApJ...472..460F}, \citealt{2006A&A...456...23B}) or for the galaxies (\citealt{2005MNRAS.358..139F}). Some works also suggest that the polarity of the bias is dependent on several factors such as the galaxy selection procedure or the physics implemented in the simulation (\citealt{Evrard:2008vo}, \citealt{2010ApJ...708.1419L}, \citealt{2013MNRAS.430.2638M}).\\
\indent The expense of obtaining spectroscopic redshifts places limitations on acquiring velocities for a complete sample of cluster members. In practical terms, luminous galaxies are often prioritised ahead of fainter ones, and photometric pre-selection may also result in the preferential selection of red over blue galaxies. The limiting magnitude of a given galaxy survey also results in a biased selection of the luminosity function. It is crucial to understand the impact of both prioritising the brighter galaxies for selection and using a fraction of member galaxies has on the measurement of cluster velocity dispersions in both simulated and observational data. In this work we extend previous analysis exploring in depth the impact of limiting magnitudes and sub-sample selections of the brightest galaxies using a simulated galaxy catalogue. We also perform similar sub-sample selection on an observational galaxy catalogue, confirming the presence of this bias in observational data. The paper is organised as follows. We describe the \cite{2007MNRAS.375....2D} semi-analytic model (SAM) galaxy catalogue followed by a description of the SDSS DR8 catalogue used for this study in Section 2. In Section 3 we present our method and results and we end with our discussion and conclusions in Section 4. Throughout the paper we adopt a $\Lambda$ cold dark matter ($\Lambda$CDM) cosmology with $\Omega_{0}=0.25$, $\Omega_{\Lambda}=0.75$, and a Hubble constant of $\rm H_{\rm 0} = 73$ $\rm kms^{\rm -1} Mpc^{\rm -1}$.\renewcommand{\tabcolsep}{0.68cm}
\begin{center}
\begin{table*}
\caption{Mass distribution of \cite{2007MNRAS.375....2D} semi-analytic clusters across five low redshift snapshots. Clusters are selected according to two criteria: $N_{\rm gal} \geq 50$ (within $\rm R_{\rm 200}$) and halo $M_{\rm 200} \geq 5$x$10^{13} M_{\rm \odot}$.}
\begin{tabular}{c | c | c | c | c}
\hline \hline
\large $z$ & \multicolumn{4}{ c }{\normalsize $\rm N_{\rm clusters}$}
 \\[0.5ex]
 & \footnotesize log $M_{\rm 200} < $ 13.9& \footnotesize 13.9 $\rm \leq$ log $M_{\rm 200} <$ 14.3 & \footnotesize 14.3 $\rm \leq$ log $M_{\rm 200} <$ 14.7 & \footnotesize 14.7 $\leq$ log $M_{\rm 200}$\\
\hline 
0.00 & 454 & 1279 & 286 & 35 \\
0.21 & 405 & 1046 & 215 & 16\\
0.41 & 370 & 862 & 129 & 7\\
0.62 & 355 & 617 & 68 & 4\\
0.83 & 288 & 434 & 38 & 1\\
\hline
\end{tabular}
\end{table*} 
\end{center}
\section{Data}
A simulated galaxy catalogue where the `true' halo mass is known is an essential tool to probe the observational limitations of constructing cluster masses based upon the spectroscopic redshifts of member galaxies. Testing these limitations across a range of cluster masses and redshifts enables any evolution of bias to be identified. Furthermore, exploring the impact of bias in observational data serves to re-enforce that a member galaxy selection procedure which distorts the measured cluster mass must be avoided. 
\subsection{Simulated data}
\label{sec:Sim Data}
We use the $N=2160^{3}$ particle Millennium Simulation (\citealt{2005Natur.435..629S}), which tracks the evolution of DM particles of mass $8.6$ x $10^{8}h^{-1} M_{\rm \odot}$ from $z=127$ to $z=0$ within a comoving box of size 500 $(\rm h^{-1} Mpc)^{3}$. The simulation adopts a flat $\Lambda$CDM cosmology with the following parameters: $\Omega_{\rm 0}=0.25$, $\Omega_{\Lambda}=0.75$, $\sigma_{8}=0.9$, $n=1$ and $h=0.73$, reflecting the values deduced by the Two-Degree Field Galaxy Redshift Survey (2dFGRS; \citealt{2001MNRAS.328.1039C}) and the Wilkinson Microwave Anisotropy Probe first year data (WMAP; \citealt{2003ApJS..148..175S}). Haloes with at least 20 particles are identified using a standard friends-of-friends (FOF) group finder (\citealt{1976ApJS...32..409T}, \citealt{1985ApJ...292..371D}) with linking length $b=0.2$, and subhaloes are then found using the SUBFIND substructure algorithm (\citealt{2001MNRAS.328..726S}).\\
\indent The \citet{2007MNRAS.375....2D} SAM of galaxy formation applied to the halo merger trees originates from \cite{1999MNRAS.303..188K}, \cite{2001MNRAS.328..726S} and \cite{2004MNRAS.349.1101D} and is a slightly modified version of that used in \cite{2005Natur.435..629S}, \cite{2006MNRAS.365...11C} and \cite{2006MNRAS.366..499D}. We refer the reader to these papers for further details of the model. Each FOF group contains one `central' galaxy that is located at the position of the most bound particle of the halo. The other `satellite' galaxies that constitute the cluster are originally central galaxies of haloes that merged to form the FOF group, maintaining the positions and velocities of the surviving core of their halo. In order to focus on more massive clusters containing sufficient numbers of galaxies for a detailed dynamical analysis, we impose a halo mass cut of $5$x$10^{13} M_{\rm \odot}$ and require that $N_{\rm gal} \geq 50$. Note that the halo mass is defined as the mass within the radius where the halo has an overdensity of 200 times the critical density of the Universe.\\
\indent We select the haloes from a range of low redshift snapshots $z = 0.00, 0.21, 0.41, 0.62, 0.83$, so that any evolution of bias with redshift is visible (see Table 1). We place an overall absolute I-band magnitude limit of $M_{\rm i} = -18.25$ to reflect the turn over of the luminosity functions constructed for each redshift snapshot, avoiding any introduction of bias due to incompleteness as a result of the resolution limit of the Millennium Simulation (\citealt{2006MNRAS.365...11C}).
\subsection{Observational data}
\label{sec:Obcentres Data}
We select cluster galaxies from the \cite{2012A&A...540A.106T} SDSS DR8 groups and clusters catalogue to reflect our intention of using an observational sample that has a similar construction in terms of the member selection, redshift and cluster $N_{\rm gal}$ as the low redshift snapshots we use from the semi-analytic \cite{2007MNRAS.375....2D} model. The catalogue adopts a lower apparent magnitude limit of $m_{\rm r} \leq $ 17.77 reflecting the magnitude to which the spectroscopic galaxy sample is complete (\citealt{2002AJ....124.1810S}). \cite{2012A&A...540A.106T} correct the redshifts of the galaxies for the motion relative to the CMB and the comoving distances are calculated by \cite{2002sgd..book.....M}, leaving a sample of 576493 galaxies with redshifts 0.009 $\leq$ $z$ $\leq$ 0.200. K-corrections using the KCORRECT (v4 2) algorithm (\citealt{2007AJ....133..734B}) were applied to the R-band absolute magnitudes and evolution corrections were made using the \cite{2003ApJ...592..819B} luminosity evolution model. The magnitudes correspond to the rest-frame at redshift $z = 0$. Further detail regarding a similar data reduction of the SDSS DR7 is described in \cite{2010A&A...514A.102T}.\\
\indent The construction of the group catalogue is based upon the FOF method whereby galaxies are linked into systems using both radial and transversal scaled linking lengths. These are chosen to reflect \cite{2012A&A...540A.106T}'s goal of obtaining groups to estimate the luminosity density field and to study properties of the galaxy network with the requirement that the group properties do not change with distance. The properties (richness, mean sizes, etc.) of the groups obtained have been assessed for any distance dependence and are found to be homogenous (\citealt{2009ApJ...696.1441T}, \citealt{2012A&A...540A.106T}).\\
\indent Out of the 77858 groups, we select clusters with $N_{\rm gal} \geq 50$ to perform our analysis ensuring that we are selecting massive clusters with sufficient numbers of tracers present in the catalogue to provide useful subsamples. Any clusters flagged with having at least one member outside the survey mask are excluded. The catalogue is also affected by fibre collisions due to the fact that the minimum separation between spectroscopic fibres is 55''. \cite{2008A&A...479..927T} have shown that this issue does not produce any strong effects when using their group-finding algorithm; they estimate the fraction of missing pairs in the catalogue to be about 8\%. Fibre collisions resulting in missing galaxies are more likely to occur in the crowded centres of clusters, which is a potential issue when focussing on brighter galaxies such as the brightest cluster galaxy (BCG), often located near the cluster centre. For this reason our analysis is carried out several times with different criteria for selecting the clusters both with and without fibre collision flags. We find that using clusters that contain fewer than three galaxies with potentially missing neighbours has little or no effect on our results. For this reason we only exclude clusters with more than two galaxies that have potentially missing neighbours. These selection criteria reduce the \cite{2012A&A...540A.106T} data to 38 clusters in the sample (with properties shown in Table 2) that are evenly distributed across the redshift range 0.021 $\leq z \leq$ 0.098. 
\begin{center} 
\begin{table*} 
\caption{Properties of the \cite{2012A&A...540A.106T} SDSS DR8 groups and clusters sample. Clusters are selected under the criteria that they must have less than two galaxies with potentially missing neighbours due to fibre collisions and $N_{\rm gal} \geq 50$.}

\begin{tabular}[ht]{c|c|c|c|c|c}
\hline \hline
\large $N_{\rm clusters}$ & Median $z$ & Min. $N_{\rm gal}$ & Max. $N_{\rm gal}$ & Min. $\sigma$ $\rm [kms^{-1}]$ & Max. $\sigma$ $\rm [kms^{-1}]$\\
\hline 
38 & 0.06 & 50 & 207 & 212.9 & 936.8 \\
\hline
\end{tabular}
\end{table*} 
\end{center}
\section{Measuring cluster velocity dispersion}
\subsection{`True' dark matter halo and total cluster galaxy velocity dispersions}
\begin{figure}
 \centering
  \includegraphics[width=0.89\textwidth]{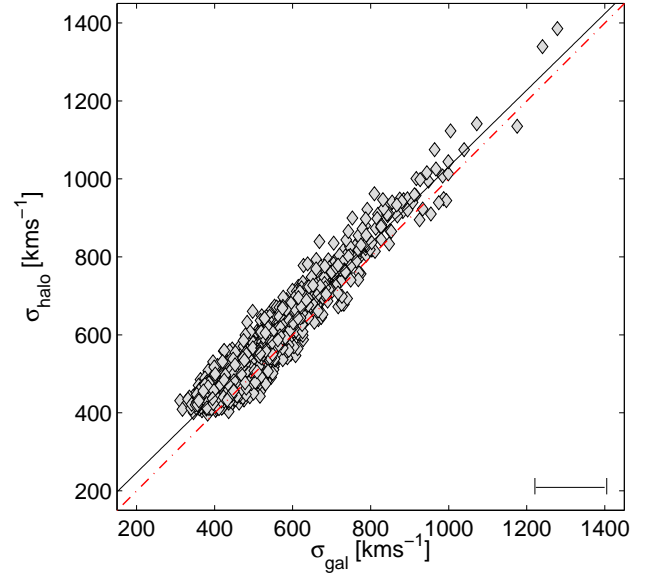}
   \caption{DM halo velocity dispersion taken from the Millennium Simulation (\citealt{2005Natur.435..629S}) catalogue against the velocity dispersion obtained from the semi-analytic galaxies (with no galaxy selection or radial cut) at $z=0$. The clusters are represented as black diamonds and the corresponding linear fit to the data with equation of the line $y=0.98x+51$ is shown as a solid black line. The red dashed line represents a 1:1 relation for comparison and the solid black horizontal line (bottom right) indicates the typical size of the bootstrap resampled $\rm 1\sigma$ error bars. The halo velocity dispersion is systematically higher than the velocity dispersion measured from the semi-analytic galaxies.}
\label{fig:DMsigmaVsGalSigma}
\end{figure}

\begin{figure}
 \centering
  \includegraphics[trim = 24mm 1mm 1mm 6mm, clip, width=0.59\textwidth]{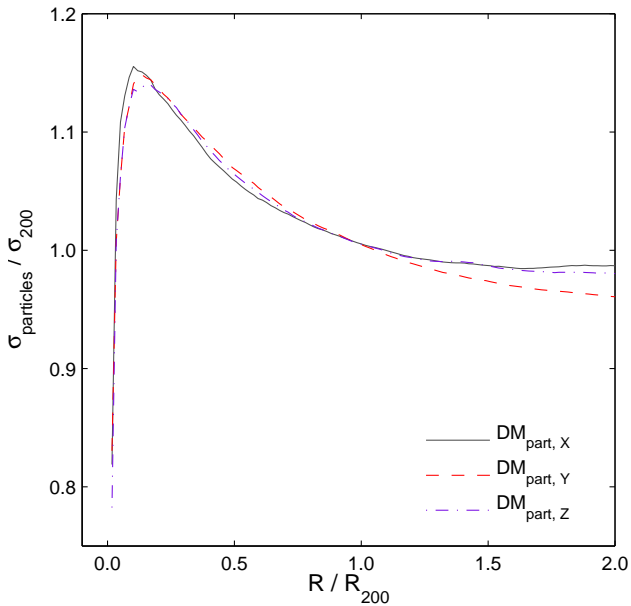}
   \caption{Velocity dispersion as a function of radius from the halo centre of 108 stacked \cite{2006MNRAS.365...11C} semi-analytic clusters. The solid black, dashed red and dot dashed purple curves signify the dispersion of the DM particles for the $x$, $y$ and $z$ lines of sight respectively. The rise and fall of the DM particle dispersion curve demonstrate that the velocity dispersion is sensitive to the radius at which it is measured.}
\label{fig:CrotonNormalisedSigmaRadial}
\end{figure}

To determine whether it is possible to recover the `true' halo mass from the \cite{2007MNRAS.375....2D} semi-analytic galaxies, we first compare the velocity dispersions calculated from all member galaxies (with no radial restriction) of clusters at z=0 to the velocity dispersions of their parent DM haloes.\\
\indent The velocity dispersions of the semi-analytic galaxy clusters are measured using the bi-weight scale estimator. This is used in place of the standard deviation, providing a more robust estimator in the case of a non-Gaussian or contaminated normal distribution (\citealt{1990AJ....100...32B}). To reduce the intrinsic scatter introduced by using a single line of sight, reported to be three times that of utilizing all three (\citealt{2012arXiv1203.5708S}), we take advantage of the three-dimensional nature of the simulation box. The cluster velocity dispersion is measured by taking the mean of the velocity dispersions measured along the $x$, $y$, and $z$ axes. Bootstrap resampling is performed for each cluster obtaining 68$\%$ asymmetric error bars. The `true' DM halo dispersion values taken from the catalogue are calculated using the standard deviation of the velocities of all particles within a spherical $\rm R_{200}$ aperture (\citealt{2005Natur.435..629S}).\\
\indent This `true' DM halo dispersion is compared with the semi-analytic galaxy cluster dispersion (using all cluster members) for haloes at $z=0$ as shown in Figure~\ref{fig:DMsigmaVsGalSigma}. Note that these semi-analytic cluster velocity dispersions are calculated with no radial truncation of the member galaxies. Clusters are represented as filled black diamonds with error bars deduced from bootstrap resampling. The black solid line is a linear fit through the data with the equation of the line $y=0.98x+51$ and the red dot dashed line shows a 1:1 relation. The halo velocity dispersion is systematically higher than the velocity dispersion measured from the galaxies with a y-intercept of 51 translating to an offset of up to $13\%$ for the lower velocity dispersion clusters.\\
\indent In order to examine this systematic difference further, we extend this analysis to a different semi-analytic catalogue for which the underlying dark matter particle properties are available to us. We compare the velocity distribution of DM particles constituting 108 DM haloes with the velocity distribution of 108 \cite{2006MNRAS.365...11C} semi-analytic galaxy clusters populated on top of these haloes. To distinguish whether there is a bias for the population as a whole, we normalise the velocities of the DM particles and the galaxies to the parent halo velocity dispersion and stack into a single distribution.\\ 
\indent The bi-weight scale estimator identifies a larger velocity dispersion for the stacked DM particle distribution than the stacked galaxy velocity distribution. A two-sample Kolmogorov-Smirnov (KS) test is employed with the null hypothesis that the stacked galaxy velocity distribution and the stacked DM particle distribution are from the same continuous distribution. The KS test delivers a p-value of 0.03 and rejects the null hypothesis at the $5\%$ significance level, indicating that even with the full semi-analytic galaxy tracer information it is difficult to recover the `true' halo velocity distribution.\\
\indent We calculate the velocity dispersion of the DM particles within different radial shells of the haloes, to examine how the velocity dispersion varies as a function of distance from the halo centre. We normalise the dispersion at different radial shells by the velocity dispersion measured using all particles within $\rm R_{\rm 200}$ and plot our results in Figure~\ref{fig:CrotonNormalisedSigmaRadial}. The solid black, dashed red and dot dashed purple curves signify the stacked dispersion of the DM particles for the $x$, $y$ and $z$ lines of sight respectively. Figure~\ref{fig:CrotonNormalisedSigmaRadial} demonstrates that the velocity dispersion is sensitive to the radius at which it is measured, with a $10\%$ difference in dispersion from 0.5 $\rm R_{\rm 200}$ - 1.0 $\rm R_{\rm 200}$. For this reason it is important that there is a consensus among the literature as to the radius at which the velocity dispersion is measured. If parameters such as the mass of the halo are typically measured at $R_{\rm 200}$, the velocity dispersion should also be measured at the same radius for consistency. For this reason, we only select galaxies within the halo $R_{\rm 200}$ in subsequent analysis. When we apply this criteria, we find that the the DM and galaxy velocity dispersion converge, as shown in Figure~\ref{fig:magnitudecutredshift0}.\\
\indent In addition to radius, the velocity dispersion is also sensitive to shape of the volume within which it is measured. For example, in the Millennium Simulation (\citealt{2005Natur.435..629S}), the DM particles are tagged as members of the halo using an FOF group finding method which, characteristically, identifies extended structures. If the velocity dispersion is measured within this FOF group, particles at the very edge of the FOF object, which are not physically bound to the halo, will contribute to the velocity dispersion.\\
\indent With no radial cut we find that the semi-analytic galaxy velocity dispersions are systematically lower than the DM halo velocity dispersions across all halo masses for both the \cite{2007MNRAS.375....2D} and \cite{2006MNRAS.365...11C} semi-analytic galaxies. This is in contrast to recent work by \cite{2013MNRAS.430.2638M} who find that the discrepancy is positive i.e. $\rm \sigma_{gal} / \sigma_{DM} \geq 1$ for higher mass systems ($h(z)M_{200} > $ 3x$10^{14}M_{\rm \odot}$) and negative i.e. $\rm \sigma_{gal} / \rm \sigma_{DM} \leq 1$ for lower mass systems ($h(z)M_{200} < 10^{14}M_{\rm \odot}$). It is noted that \cite{2013MNRAS.430.2638M} find that the polarity of the velocity bias not only depends on whether or not the simulation includes baryonic physics but how the baryonic physics is implemented, a potential origin of the difference in results. It is unlikely that this disparity is a result of a difference in the luminosity or mass cuts applied in the two samples. We apply a luminosity cut of -18.25, which translates to a luminosity cut of $\rm \sim 8.7x10^{8} L_{\rm \odot}$ and \cite{2013MNRAS.430.2638M} apply a stellar mass cut of $\rm M_{\rm stellar} = 3x10^{9} M_{\rm \odot}$. The ratio of the two cuts is $\rm M_{stellar}/L \sim 3.4$ which is reasonable for a galaxy observed in the I-band, indicating that this is not the source of the disparity.


\subsection{Selecting galaxy tracers by absolute magnitude cut}
We want to identify whether the presence of a limiting magnitude depth in a galaxy survey has a significant effect on the measured velocity dispersion of a cluster. To do this, the velocity dispersion of each semi-analytic cluster at $z=0$ is calculated using only the member galaxies within $\rm R_{\rm 200}$ that have magnitudes below a given absolute I-band magnitude, imposing increasingly faint magnitude limits. These I-band magnitude values are taken from $M_{\rm i} = -24.0$ to $M_{\rm i} = -18.5$.\\
\indent The velocity dispersions are calculated at each magnitude limit using one of three estimators: the bi-weight scale estimator, the gapper estimator and the standard deviation. As recommended by \citet{1993ApJ...404...38G} and \citet{1990AJ....100...32B}, the bi-weight scale estimator is employed if the number of galaxies remaining in the sample $N^{i}_{gal}$, after the magnitude limit is $N^{i}_{gal} > 15$. The gapper estimator is used if $5 \leq N^{i}_{gal} \leq 15$ and if $N^{i}_{gal} \leq 5$, the standard deviation is used. The velocity dispersions are then normalised by the parent DM halo velocity dispersion.\\
\begin{figure}
  \includegraphics[trim = 27mm 1mm 1mm 6mm, clip, width=0.58\textwidth]{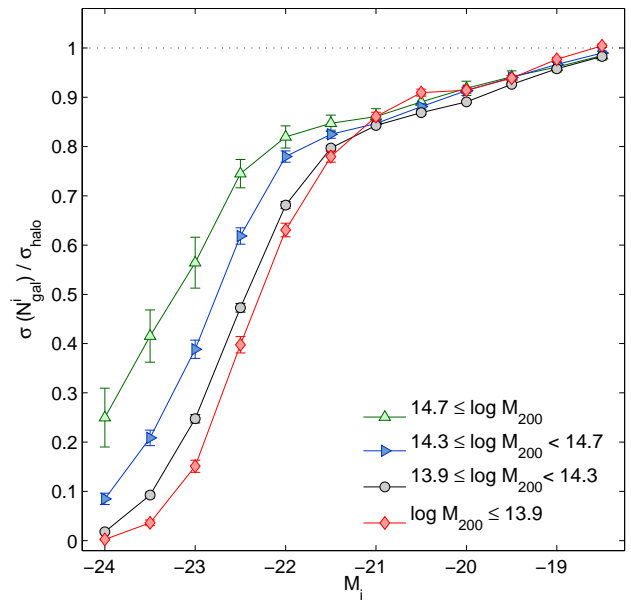}
   \caption{The evolution of the velocity dispersion of the semi-analytic clusters at $z=0$ with brighter absolute I-band magnitude limits excluding fainter galaxies. Note that member galaxies are selected within the halo $R_{\rm 200}$. The velocity dispersion is normalised by the halo velocity dispersion and the clusters are binned according to halo $M_{200}^{\rm Crit}$. The error bars signify the standard error on the mean of clusters included in each cluster mass bin. A brighter adopted I-band magnitude limit leads to an increasingly underestimated cluster velocity dispersion, a result that is more severe for clusters with lower halo mass.}
\label{fig:magnitudecutredshift0}
\end{figure}
\indent We show the velocity dispersion as a function of imposed I-band magnitude limits in Figure~\ref{fig:magnitudecutredshift0}. The clusters are binned according to halo $M_{200}^{\rm Crit}$ (the mass within the radius where the halo has an overdensity 200 times the critical density of the Universe), with bin sizes chosen to ensure that there are enough clusters in the higher mass bins to perform statistical analysis. See Table 1 for the distribution of clusters in each mass bin for each of the five low redshift snapshots. The error bars signify the standard error on the mean of clusters included in each cluster mass bin. \\
\indent The figure identifies that a brighter absolute I-band magnitude cut results in significant underestimation of the cluster velocity dispersion. This underestimation is most severe for the lower halo mass clusters with velocity dispersions underestimated on the order of $10-40\%$ more than clusters in the highest halo mass bin at an absolute I-band magnitude cut of $M_{\rm i} \leq -21$. This mass-dependent underestimation is evident across all five redshift snapshots. It is also important to note that the clusters for which a velocity dispersion cannot be calculated due to lack of cluster members are not included in the mean values hence the level of underestimation is the `best-case' scenario.\\
\indent In practical terms, the size of the underestimation in velocity dispersion is dependent upon the limiting magnitude of a given survey. To put the absolute magnitude limits applied in this analysis into context, the SDSS survey apparent R-band magnitude limit of $m_{\rm r} \leq $ 17.77 corresponds to an absolute R-band magnitude of $M_{\rm r} \leq $ -17.21 for the lowest redshift cluster in our sample. For the highest redshift cluster this corresponds to $M_{\rm r} \leq $ -20.60. Assuming a typical $r-i$ colour of an elliptical cluster galaxy as $r-i=0.4$ (\citealt{2006MNRAS.366..717C}), this corresponds to an absolute I-band magnitude of $M_{\rm i} \leq $ -17.61 and $M_{\rm i} \leq $ -21.00 for the lowest and highest redshift clusters respectively.
It is clear from Figure~\ref{fig:magnitudecutredshift0} that this constraint has the potential to severely negatively bias the measured velocity dispersion, and hence dynamical masses of clusters, which in turn will affect any cosmological information they provide.


\subsection{Selecting galaxy tracers by $\rm N_{\rm gal}$ cut}
\begin{figure}
 \centering
  \includegraphics[trim = 25mm 1mm 1mm 6mm, clip, width=0.60\textwidth]{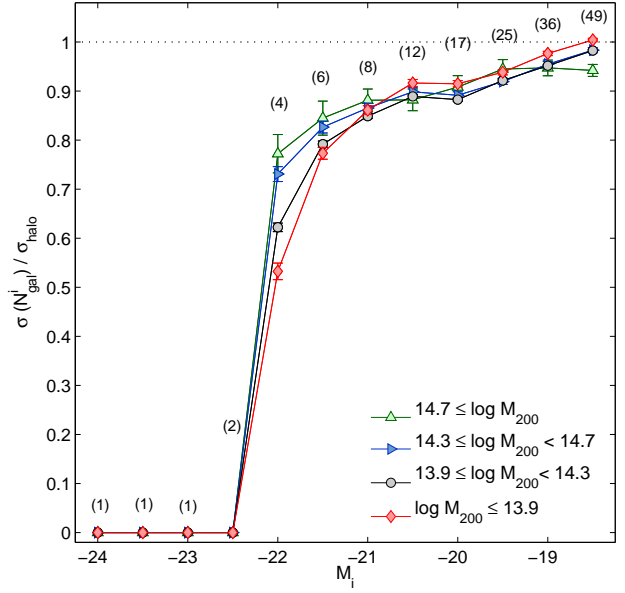}
   \caption{The velocity dispersion of clusters at $z=0$ with both an increasing I-band absolute magnitude limit and a number selection limit. The number of galaxies in clusters are limited to the mean number of galaxies in clusters with halo $M_{200}^{\rm Crit}$ in the lowest mass bin (within the halo $R_{\rm 200}$). These galaxies are randomly selected from the galaxies about the imposed I-band absolute magnitude limit. The error bars signify the standard error on the mean of clusters included each cluster mass bin. The underestimation in the velocity dispersion is more severe for the lower mass clusters with both a magnitude limit and a number limit imposed.}
\label{fig:MagnitudeCutAndLimitedNgalGalSigmaMsSigma0}
\end{figure}

\begin{figure}
 \centering
  \includegraphics[trim = 22.5mm 1mm 1mm 6mm, clip, width=0.59\textwidth]{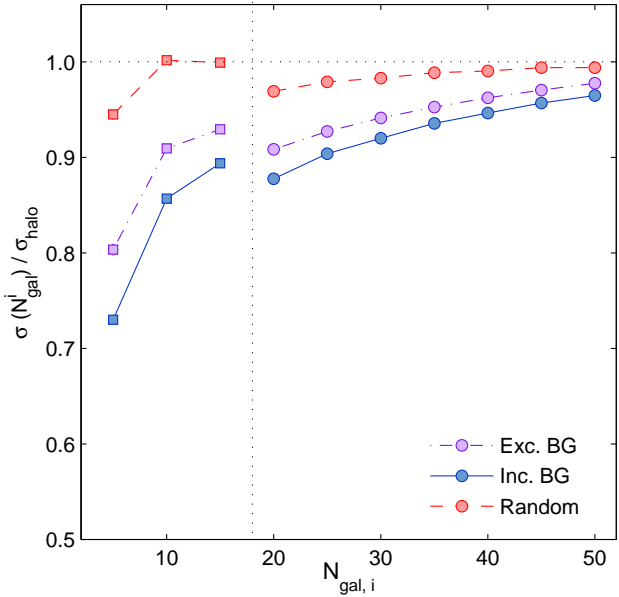}
   \caption{The velocity dispersion of clusters at $z=0$ with the inclusion of the $N_{\rm gal}$ brightest galaxies in the cluster both excluding and including the brightest galaxy. The purple dot dashed and blue solid lines represent the mean of the normalised velocity dispersion when taking the brightest $N_{\rm gal}$ galaxies with the central galaxy excluded and included respectively (and with all member galaxies within the halo $R_{\rm 200}$). The red dashed line represents the mean of the normalised velocity dispersion for all clusters when taking the $N_{\rm gal}$ galaxies randomly. The circle markers represent instances where the bi-weight scale estimator is employed and the square markers represent instances where the standard deviation and the gapper estimator are used (standard deviation for $N_{\rm gal}=5$ and the gapper estimator for $N_{\rm gal} = 10,15$). Excluding the brightest galaxy reduces the brighter galaxy bias by $2-6\%$.}
\label{fig:central_vs_no_central}
\end{figure}

When measuring dynamical cluster properties such as the `true' cluster velocity dispersion and dynamical substructure, there is a consensus among the literature that obtaining on the order of 20 cluster members is sufficient (e.g., \citealt{1990AJ....100...32B}, \citealt{2010A&A...521A..28A} and \citealt{2012MNRAS.421.3594H}). In addition to exploring the impact of different survey magnitude limits, we now examine how imposing a number cut changes the recovered velocity dispersion, reflecting the (commonly assumed) spectroscopic strategy when restricted to a subsample of member galaxies.\\ 
\begin{figure}
 \centering
  \includegraphics[trim = 27mm 1mm 1mm 6mm, clip, width=0.58\textwidth]{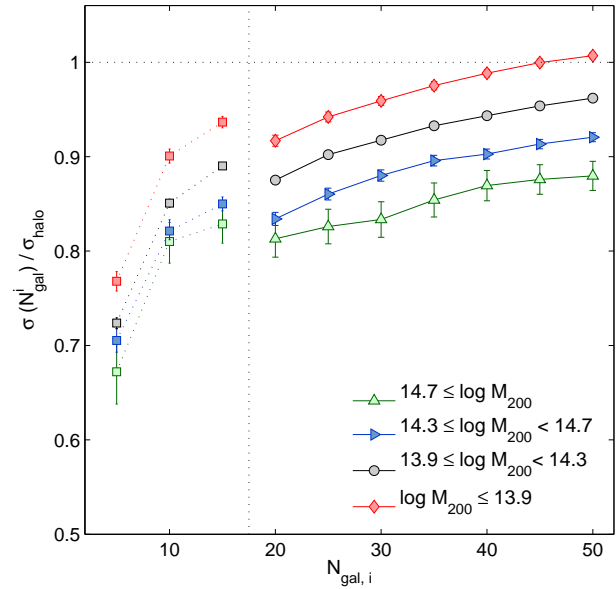}
   \caption{The velocity dispersion of clusters at $z=0$ with the inclusion of the $N_{\rm gal}$ brightest galaxies in the cluster (within the halo $R_{\rm 200}$). The velocity dispersion is normalised by halo velocity dispersion and clusters are binned according to halo $M_{200}^{\rm Crit}$. The curves to the right of the vertical dotted line represent instances where the bi-weight scale estimator is employed and the square markers to the left of the vertical dotted line represent instances where the standard deviation and the gapper estimator are used (standard deviation for $N_{\rm gal}=5$ and the gapper estimator for $N_{\rm gal} = 10,15$). The cluster velocity dispersion is increasingly underestimated with the inclusion of only the brightest $N_{\rm gal}$ selection of galaxies. In contrast to Figure~\ref{fig:magnitudecutredshift0}, the higher mass clusters are the most severely affected.}
\label{fig:NgalBrightestRedshift0}
\end{figure}
\indent First we examine the selection of the $N_{\rm gal}$ brightest cluster galaxies, an approach often adopted in the case of limited spectral fibres or mask slits. For each semi-analytic cluster we rank member galaxies according to their absolute I-band magnitude and calculate the velocity dispersion using the selections of the brightest galaxies from $N_{\rm gal} = 5$ to $N_{\rm gal} = 50$. These velocity dispersions are then normalised to the DM halo velocity dispersion.\\
\begin{figure}
 \centering
  \includegraphics[trim = 28mm 1mm 1mm 6mm, clip, width=0.59\textwidth]{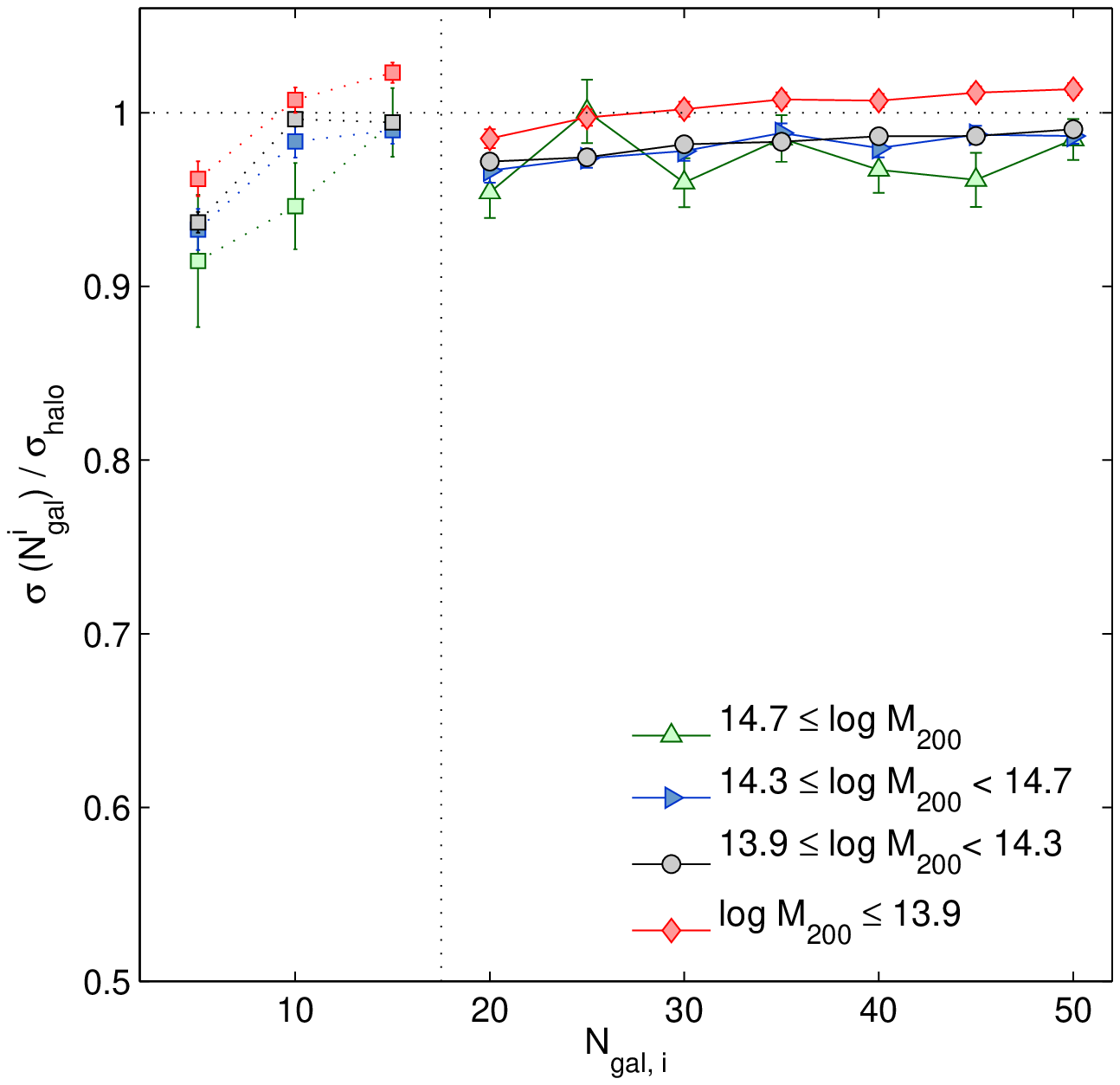}
   \caption{The velocity dispersion of semi-analytic clusters at $z=0$ with the inclusion of a randomly selected $N_{\rm gal}$ galaxies in the cluster (within the halo $R_{\rm 200}$). The velocity dispersion is normalised by halo velocity dispersion and clusters are binned according to halo $M_{200}^{\rm Crit}$. The curves to the right of the vertical dotted line represent instances where the bi-weight scale estimator is employed and the square markers to the left of the vertical dotted line represent instances where the standard deviation and the gapper estimator are used (standard deviation for $N_{\rm gal}=5$ and the gapper estimator for $N_{\rm gal} = 10,15$). The error bars signify the standard error on the mean of clusters included each cluster mass bin. The velocity dispersion remains fairly constant with the increasing number of galaxies selected randomly for velocity dispersion measurement.}
\label{fig:NgalRandomRedshift0}
\end{figure}
\begin{figure*}
 \centering
  \includegraphics[trim = 30mm 0mm 2mm 2mm, clip, width=1.15\textwidth]{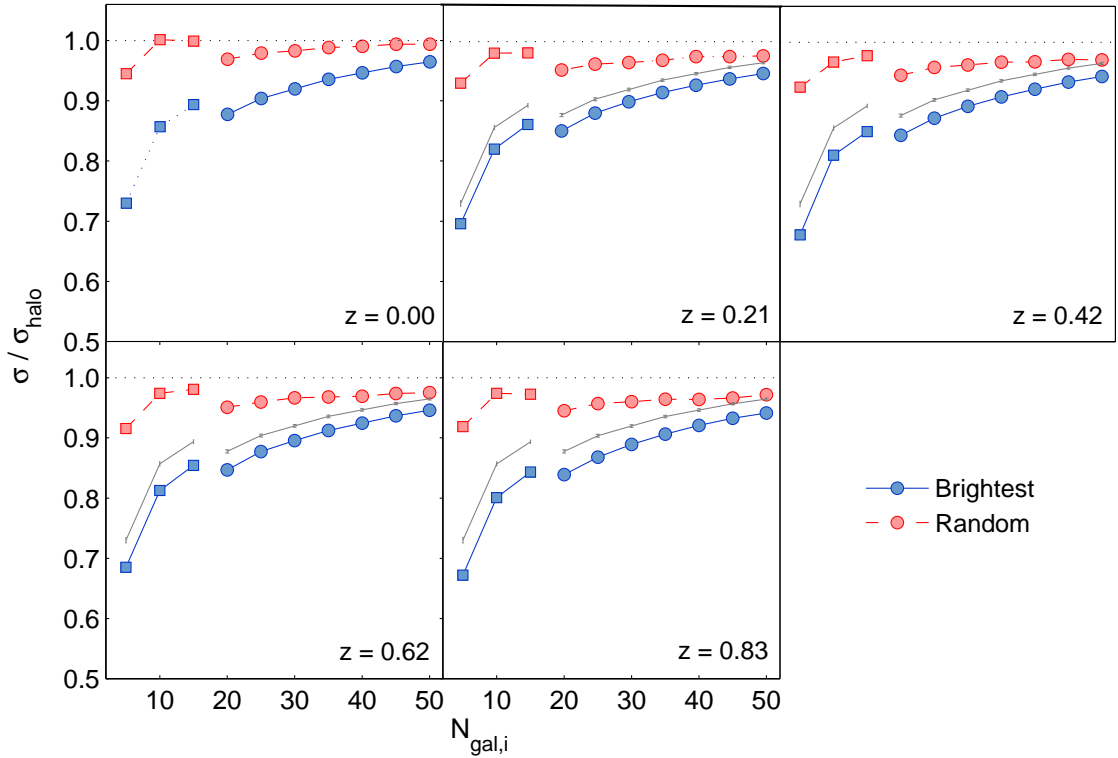}
   \caption{The evolution of the velocity dispersion measured when taking the $N_{\rm gal}$ brightest galaxies and randomly selecting $N_{\rm gal}$ galaxies at $z=0.00, 0.21, 0.41, 0.62$ and $0.83$ (within the halo $R_{\rm 200}$). The blue solid and red dashed curves refer to the mean normalised velocity dispersion of all clusters using the $N_{\rm gal}$ brightest selection and random selection respectively. The $z=0$ brightest selection curves are plotted in grey on the higher redshift subplots to highlight the redshift evolution of this systematic bias. The circle markers represent instances where the bi-weight scale estimator is employed and the square markers represent instances where the standard deviation and the gapper estimator are used (standard deviation for $N_{\rm gal}=5$ and the gapper estimator for $N_{\rm gal} = 10,15$). Measuring the velocity dispersion with a brighter selection of galaxies introduces a bias on the order $5-15\%$ more than when taking a random selection of galaxies across all redshift snapshots.}
\label{fig:NgalBrightestRandomAllRedshifts}
\end{figure*}
\indent We plot the velocity dispersion as a function of the $N_{\rm gal}$ brightest galaxy selections in Figure~\ref{fig:NgalBrightestRedshift0}. The difference between the mean normalised velocity dispersion at low and high $N_{\rm gal}$ selections demonstrate that the cluster velocity dispersion is increasingly underestimated with the inclusion of only the $N_{\rm gal}$ brightest galaxies. It is also clear from the difference in the distribution of the four mass bin curves that taking the $N_{\rm gal}$ brightest cluster members results in a more severe bias for the higher mass clusters on the order of $10-15\%$ more than the lower mass clusters.\\
\indent In order to differentiate whether the extent of the underestimation is simply due to a lack of galaxy tracers or whether it is the type of galaxies that are being excluded, we repeat the process above but with $N_{\rm gal}$ selections of galaxies chosen randomly with respect to their absolute I-band magnitude. The velocity dispersion is calculated using selections of the galaxies from $N_{\rm gal} = 5$ to $N_{\rm gal} = 50$ and the resulting normalised and mass binned values are plotted in Figure~\ref{fig:NgalRandomRedshift0}. The mass dependence apparent in Figure~\ref{fig:NgalBrightestRedshift0} reduces dramatically when using a random $N_{\rm gal}$ selection of galaxies at $z=0$.\\
\indent The reversal of the mass dependence from Figure~\ref{fig:magnitudecutredshift0} to Figure~\ref{fig:NgalBrightestRedshift0} could be a consequence of the proportion of member galaxies that are included in the two methods. In the case of imposing increasingly brighter magnitude limits, the proportion of galaxies that are excluded is greater for the lower mass haloes, leaving only the brighter galaxies (with less `dilution' from the fainter galaxies) which may cause a more severe underestimation. However when we do not impose a magnitude cut but only select the brightest $N_{gal}$ galaxies, we are maintaining the same number of cluster galaxies across all halo masses. The brightest $N_{gal}$ galaxies of the higher mass haloes are a more biased subsample than that of the lower mass haloes as the higher mass haloes have, on average, more member galaxies. For the lower mass haloes, with fewer member galaxies, the brightest $N_{gal}$ members will include a higher proportion of fainter galaxies.\\
\indent When we compare the mean normalised velocity dispersion of the whole cluster population when only selecting either the top $N_{\rm gal}$ brightest galaxies or a random selection of $N_{\rm gal}$ galaxies at each redshift uncovers a distinct difference in the two selection methods, as is evident in Figure~\ref{fig:NgalBrightestRandomAllRedshifts}. In all subplots the blue solid and red dashed curves refer to the mean normalised velocity dispersion of all clusters using the $N_{\rm gal}$ brightest selection and random selection respectively. The $z=0$ curves are plotted in grey on the higher redshift subplots to highlight the redshift evolution of this systematic bias. There is a $5-20\%$ underestimation when using on the brighter galaxies to estimate the velocity dispersion indicating that dynamical friction of the brighter galaxies has a significant impact on the measured velocity dispersion. In contrast to \cite{2012arXiv1203.5708S}, who find no redshift dependence of this bias, we find that, for both the brightest $N_{\rm gal}$ and randomly chosen $N_{\rm gal}$ selections, the bias increases with redshift, underestimating the cluster velocity dispersion by a further factor of up to $5\%$ from $z=0$ to $z=0.83$. Though the polarity of the ratio between the semi-analytic and dark matter halo dispersions is the opposite of that found by \cite{2013MNRAS.430.2638M}, as discussed in Section 3.1, they also find that this ratio increases with redshift.\\
\indent It is also important to understand how the bias changes with cluster mass in the commonly occurring case of both a limiting magnitude and a restricted number of galaxy redshifts. To do this we combine the two approaches described above and first impose varying I-band magnitude limits taken from $M_{\rm i} = -24.0$ to $M_{\rm i} = -18.5$. We then calculate the mean number of galaxies remaining after each magnitude limit cut for the lowest mass cluster bin. We use this mean number of galaxies in the same $N_{\rm gal}$ procedure as described above to select $N_{\rm gal}$ random galaxies to calculate the velocity dispersion. Again, we plot the normalised mean velocity dispersion values binned by cluster mass in Figure~\ref{fig:MagnitudeCutAndLimitedNgalGalSigmaMsSigma0}. This figure demonstrates that the dispersions of the higher mass haloes are substantially more underestimated than their lower mass counterparts above $M_{\rm i} = -21.0$, in contrast to Figure~\ref{fig:magnitudecutredshift0} when only imposing absolute I-band magnitude limits. If the magnitude limit is fainter than $M_{\rm i} = -21.0$, we find little variation with mass, suggesting that clusters across the whole mass range in our sample are biased to the same extent above $M_{\rm i} = -21.0$.\\
\indent In order to determine the extent to which the brightest galaxy in the cluster induces the $5-20\%$ underestimation, we repeat our analysis but we exclude the brightest galaxy.
The purple dot dashed line in Figure~\ref{fig:central_vs_no_central} represents the mean of the normalised velocity dispersion for all clusters at $z=0$ when taking the brightest $N_{\rm gal}$ galaxies with the central galaxy excluded. The blue solid line also represents the mean of the normalised velocity dispersion for all clusters when taking the brightest $N_{\rm gal}$ galaxies but with the central galaxy included (as shown in the upper left subplot of Figure~\ref{fig:NgalBrightestRandomAllRedshifts}). The red dashed line represents the mean of the normalised velocity dispersion for all clusters when taking the $N_{\rm gal}$ galaxies randomly. We find that excluding the brightest galaxy reduces the underestimation on the order of $2-6\%$. This implies that, though the brightest galaxy does have some impact, it does not account for the majority of the underestimation.\\
\indent We find that when employing different tracer selection criteria the mass dependence of the bias changes. When solely imposing magnitude limits, the lower mass haloes are more severely affected, in contrast to the case of an $N_{\rm gal}$ brightest galaxy selection, where the velocity dispersions of the higher mass haloes are more underestimated. When taking a random selection of galaxies from all clusters the mass-dependence is reduced, however the higher mass haloes appear slightly more affected than the lower mass haloes. In the case of both a limiting magnitude and a restricted number of galaxy redshifts it is the lower mass haloes that are more affected.

\subsection{Velocity dispersions from observational data}
Having identified a clear underestimation in the measured velocity dispersion when imposing spectroscopic survey limitations such as magnitude depth and number of tracer restrictions in simulated data, we now aim to establish whether this bias is measurable in an observational sample. 
Using the CMB-corrected cluster redshift $z$, computed by \cite{2012A&A...540A.106T} as the average over all galaxies belonging to the cluster, the peculiar velocity of a member galaxy with redshift $z$ is given by:\\
\begin{equation} \nu^{\rm rest}_{\rm pec} = \frac{c(z-z_{\rm cl})}{(1+z_{\rm cl})}, \indent (v^{\rm rest}_{\rm pec} \ll c). \end{equation}\\
This is the peculiar velocity in the rest-frame of the cluster (e.g., \citealt{1979ApJ...232...18H}, \citealt{1996ApJ...462...32C}). Using the peculiar velocities, the cluster rest-frame velocity dispersion $\sigma_{\rm cl}$ is calculated using the bi-weight scale estimator as described above. For each cluster the member galaxies are ranked according to their absolute R-band magnitude (reflecting the band that the catalogue is originally selected in) and the one-dimensional velocity dispersion is measured using selections of the brightest galaxies from $N_{\rm gal} = 5$ to $N_{\rm gal} = 50$. In the absence of knowledge of the `true' DM halo velocity dispersion (information we have for the simulated data) we normalise the resultant velocity dispersions to the final $\sigma_{\rm v}$ obtained using all available cluster members. This process is repeated with galaxies selected randomly with respect to their absolute R-band luminosity to ascertain whether it is the number of galaxies used to measure the dispersion or the luminosity of the galaxies selected that is important for recovering the cluster dispersion. Note that the velocity dispersion here is measured using one line-of-sight whereas the simulated cluster velocity dispersions are measured by taking the mean of the three $x$, $y$ and $z$ lines of sight reducing the scatter in the simulated cluster dispersion values. In addition, there is no radial restriction applied to these clusters as cluster membership is defined using a transversal FOF method as discussed in Section 2.2.\\
\begin{figure}
  \includegraphics[trim = 22.5mm 1mm 1mm 6mm, clip, width=0.595\textwidth]{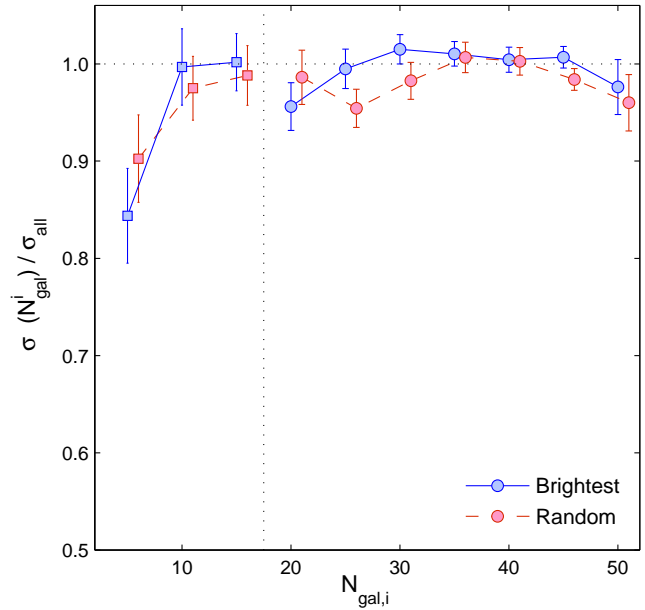}
   \caption{The mean velocity dispersion of 38 SDSS DR8 clusters measured by taking the $N_{\rm gal}$ brightest galaxies versus randomly selecting $N_{\rm gal}$ galaxies. The blue solid and red dashed curves refer to the mean normalised velocity dispersion of all clusters using the $N_{\rm gal}$ brightest selection and random selection respectively. The circle markers represent instances where the bi-weight scale estimator is employed and the square markers represent instances where the standard deviation and the gapper estimator are used (standard deviation for $N_{\rm gal}=5$ and the gapper estimator for $N_{\rm gal} = 10,15$). Error bars signify the standard error of these values and for the purpose of clarity, the red diamond markers are shifted +1 $N_{\rm gal}$ along the x-axis. The horizontal black dotted line represents the line expected if the velocity dispersion (using all member galaxies in the catalogue) is recovered.}
\label{fig:SDSSNgal}
\end{figure}
\indent The underestimation of the measured velocity dispersion when selecting only the brighter galaxies using the semi-analytic clusters is not evident in the SDSS DR8 38 cluster sample as we show in Figure~\ref{fig:SDSSNgal}. The blue curve represents the mean of the normalised one-dimensional velocity dispersion measurements of all the clusters at different $N_{\rm gal}$ selections of the most luminous absolute R-band magnitude galaxies. The red curve with also represents the mean of the normalised one dimensional velocity dispersions of all the clusters but with galaxy selections taken randomly as opposed to taken according to their absolute R-band magnitude. Error bars signify the standard error of these values and for the purpose of clarity, the random selection curve is shifted +1 $N_{\rm gal}$ along the $x-$axis.\\
\indent The velocity dispersions appear to converge to the velocity dispersions obtained using all cluster members for both selection methods for $N_{gal} \geq 5$. The difference in the velocity dispersions obtained using the two selection criteria in Figure~\ref{fig:NgalBrightestRandomAllRedshifts} for the semi-analytic data is nots discernible in Figure~\ref{fig:SDSSNgal}. This is possibly a result of the way that these SDSS galaxies are selected. A caveat of performing these two selection tests on the observational data is that the type and number of galaxies in the catalogue are influenced by the nature of the survey, making it difficult to fully assess the extent of the bias as we are unable to take a truly random sample from all member galaxies.\\
\indent Furthermore, due to the large scatter of the size of the bias on a cluster-by-cluster basis, this systematic bias would not necessarily be apparent with either a single cluster or small sample of clusters as we have here. In addition, the observational data may include interloping galaxies that are not present in the simulated data. The impact of interlopers on velocity dispersion estimates remains controversial with studies finding interlopers increase cluster velocity dispersions (\citealt{2007A&A...466..437W}) and others finding they reduce cluster velocity dispersion estimates (\citealt{1997ApJ...485...39C}, \citealt{2006A&A...456...23B}).


\section{Discussion and summary}

Using 9550 clusters across five low-redshift snapshots from the \cite{2007MNRAS.375....2D} SAM alongside the Millennium Simulation (\citealt{2005Natur.435..629S}) and 38 clusters from the \citet{2012A&A...540A.106T} SDSS DR8 groups and clusters catalogue we investigate the velocity dispersion bias from the brighter cluster galaxies. We employ different selection criteria: adopting direct magnitude cuts, $N_{\rm gal}$ brightest and $N_{\rm gal}$ random selections when calculating the cluster velocity dispersion and relate this to the parent DM halo velocity dispersion. We also compare the velocity distribution of member galaxies from 108 stacked \cite{2006MNRAS.365...11C} semi-analytic clusters with their associated DM particle velocity distribution.\\
\indent Our results demonstrate that the velocity dispersion obtained from the semi-analytic cluster galaxies is systematically lower than the parent halo dispersion on the order of $5-10\%$. The size of this discrepancy is in agreement with the literature however we find, in contrast to \cite{2013MNRAS.430.2638M}, that the bias is negative i.e. $\rm \sigma_{\rm gal} / \rm \sigma_{\rm DM} \leq 1$ for high mass clusters. We attribute both the size of the discrepancy and its polarity to several factors, the first a result of employing different definitions of the galaxy and halo velocity dispersions. We note the importance of measuring the velocity dispersion in a consistent manner within simulations and in a manner that matches the measurement of other halo properties e.g., it is measured at the same radial distance from the cluster centre as the mass ($M_{\rm 200}$). The second factor is that, for the SAM employed here (and commonly) the galaxies are populated on a biased subset of dark matter particles that are more likely to experience physical processes such as dynamical friction, leading to an underestimation of the velocity dispersion. Thirdly, \cite{2013MNRAS.430.2638M} find that the baryonic physics implemented in the simulation characterises the polarity of the velocity bias between the velocity dispersions measured using the DM and subhaloes or galaxies. All three of these mechanisms may contribute to the variation found amongst the literature.\\
\indent Furthermore, it is likely that differences in the procedure of populating the haloes with galaxies will alter how different galaxy selections affect the measured cluster dispersion. In this paper we focus on SAMs, however another common method for producing galaxy catalogues for use in large surveys is to use Halo Occupation Distribution (HOD) models (e.g., \citealt{2006MNRAS.369...68S}). We expect that the brighter galaxy bias will not be visible in HOD models due to the fact that the galaxies are essentially indistinguishable from the DM particles, as satellite galaxies are commonly distributed randomly around the centre of the halo following a spherical Navarro, Frenk \& White profile (\citealt{1996ApJ...462..563N}). In HOD models, cluster galaxy velocities are typically obtained by the sum of the velocity of the parent halo and a value of virial motion that is drawn from a Maxwell-Boltzmann distribution with dispersion that is halo mass-dependent (e.g., \citealt{2006MNRAS.369...68S}). This approach is disparate to SAMs whereby member galaxies are chosen via the tracking of subhaloes and velocities are specified by the velocity of that subhalo, maintaining more of the dynamical history of the cluster. This highlights the importance of revealing the effects of implicit assumptions made by different models; especially in the case of using galaxy mocks for observational calibration of large surveys.\\
\indent For clusters with $N_{\rm gal} \geq 50$, when calculating the velocity dispersion with $N_{\rm gal} \leq 50$ of the most luminous galaxies, the velocity dispersion is substantially underestimated. This finding is consistent with previous analysis by \cite{1992ApJ...396...35B}, \cite{2005MNRAS.359.1415G} and \cite{2012arXiv1203.5708S} who postulate that the brighter i.e., more massive galaxies are subject to more dynamical friction. In agreement with \cite{2012MNRAS.421.3594H}, we find velocity dispersions measured using $N_{\rm gal} \geq 20$ asymptote to the velocity dispersion measured when using all cluster members. In addition, we find a small redshift dependence with the velocity dispersion underestimated more, on average, as redshift increases from $z=0$ to $z=0.83$.\\
\indent In contrast to \cite{2012arXiv1203.5708S} we find a mass dependence on the underestimation of velocity dispersions obtained when selecting the $N_{\rm gal}$ brightest galaxies for analysis. The higher mass clusters are the most severely affected by choosing $N_{\rm gal}$ brightest galaxies, a dependence that is not visible when choosing random $N_{\rm gal}$ selections. This indicates that the brighter members in these high mass clusters are more affected by dynamical friction than the brighter members in the lower mass clusters.\\
\indent The underestimation of the velocity dispersion when taking $N_{\rm gal}$ selections of galaxies is not discernible in our sample of 38 clusters from the \cite{2012A&A...540A.106T} SDSS DR8 groups and clusters catalogue. In comparison to the SAM catalogue, we do not find a distinct difference when taking the brightest $N_{\rm gal}$ and random $N_{\rm gal}$ selections. However, we note that a full assessment of the brighter galaxy bias is not possible as we are unable to take a truly random sample from all member galaxies. In addition, the large scatter of the size of the bias observed on cluster-by-cluster basis emphasises that this is a systematic bias across the cluster population and is not necessarily likely to be observed with a small sample of clusters.\\
\indent In this paper we have examined the effects of selecting tracer galaxies by absolute magnitude cut, ranked luminosity number cut, or random number cut on the measured velocity dispersion. Future work will incorporate more realistic observational constraints, including the effects of fibre collisions, interlopers, and colour selection of target galaxies. Based upon the results of this paper we make a recommendation that in the realistic case of limited availability of spectral observations that a strictly magnitude-limited sampling be avoided to reduce the bias on the estimate of the velocity dispersion.\\
\section*{Acknowledgments}
We would like to thank Kathy Romer, Will Hartley, Alfonso Aragon-Salamanca, Michael Merrifield and Pascal Elahi for useful discussions which are highly appreciated. We would also like to thank the anonymous referee for constructive comments which contributed to the improvement of this paper. We would like to acknowledge funding from the
Science and Technology Facilities Council (STFC). MEG was supported by an STFC Advanced Fellowship.
\bibliographystyle{mn2e}
\bibliography{BibliographyBrighterGalaxyBias}

\label{lastpage}
\end{document}